\documentclass[prd,twocolumn,amsmath,amssymb,showpacs,floatfix,superscriptaddress,nofootinbib,preprintnumbers]{revtex4}
\usepackage{graphicx}
\usepackage{psfrag}
\usepackage{epsfig}
\usepackage{bm}
\usepackage{amsfonts}
\usepackage{float}
\usepackage{color}
\usepackage{comment}

\usepackage{amsmath,amssymb}

\def\gev{{\rm \,Ge\kern-0.125em V}}

\def\be{\begin{equation}}
\def\ee{\end{equation}}
\def\bea{\begin{equationarray}}
\def\eea{\end{equationarray}}
\def\half{\frac{1}{2}}
\def\calH{{\mathcal{H}}}
\newcommand{\lsim}{\mbox{\raisebox{-.6ex}{~$\stackrel{<}{\sim}$~}}}
\newcommand{\gsim}{\mbox{\raisebox{-.6ex}{~$\stackrel{>}{\sim}$~}}}

\begin{document}

\title{
Thawing quintessence from the inflationary epoch 
to today}

\date{\today}
\author{Gaveshna Gupta }
\email{gaveshna@prl.res.in} \affiliation{Physical Research Laboratory, Navrangpura,
Ahmedabad, 380009, India}

\author{Raghavan Rangarajan}
\email{raghavan@prl.res.in}
\affiliation{Physical Research Laboratory, Navrangpura,
Ahmedabad, 380009, India}

\author{Anjan A. Sen}
\email{aasen@jmi.ac.in} \affiliation{Centre for Theoretical
Physics, Jamia Millia Islamia, New Delhi 110025, India}

\begin{abstract}

Using the latest observational data
we obtain a lower bound on the initial value  
of the quintessence field in thawing quintessence models of dark energy.  For potentials of the form
$V(\phi)\sim\phi^{\pm2}$ we find that the initial value $|\phi_i|>7\times10^{18}\gev$. 
We then relate $\phi_i$ to the duration of inflation
by assuming that the initial value of the quintessence field is determined by quantum fluctuations of the quintessence field during
inflation.  From the lower bound on $\phi_i$ we obtain a lower bound on the number of e-foldings of inflation, namely, $N>2\times10^{11}$.
We obtain similar bounds for other power law 
potentials for which too we obtain $|\phi_{i}|>O(M_{P})$.

\end{abstract}
\maketitle

\section{Introduction}
\noindent

Cosmological observations \cite{Riess:1998cb,Perlmutter:1998np,Tonry:2003zg} 
of the past one and a half decades indicate that the
Universe is undergoing accelerated expansion. Although a non-zero
cosmological constant can explain the current acceleration of the
Universe, one still has to explain why it is so small and why only at
recent times it has started to dominate the energy density of the
Universe \cite{Zlatev:1998tr}. These issues have motivated the exploration of alternative
theories to explain the late time acceleration as due to a source 
of energy referred to as dark
energy \cite{Copeland:2006wr,Li:2011sd}. A quintessence model is one amongst such theories where the dark
energy arises from a scalar field $\phi$ rolling slowly down a potential.
 
The equation of state parameter $w$ can be defined as the ratio of the pressure to 
the energy density
\be\label{w} w = p/\rho.
\ee
A cosmological constant is equivalent to $w = -1$ whereas a quintessence
field generates a time dependent equation of state $w(t) > -1$. Caldwell
and Linder \cite{Caldwell:2005tm} showed that the quintessence models 
in which the scalar field rolls down its potential towards a  minimum can be classified into
two categories, namely $\it{freezing}$ and $\it{thawing}$ models, with quite different behavior.
In thawing models at early times the field gets locked at a value away from
the minimum of the potential  due to large Hubble damping.  At late times when Hubble damping diminishes, 
the field starts to roll down
towards the minimum. These models have a value of $w$ which begins near
$-1$ and gradually increases with time.
In freezing models 
the field rolls towards its potential minimum initially and slows down at late times 
as it comes to dominate the Universe. These models have a value of $w$
which decreases with time. In both cases $w$ $\approx -1$ around the present epoch.
Thawing models with a nearly flat
potential provide a natural way to produce
a value of $w$ that stays close to, but not exactly equal to $-1$.
The field begins with $w \approx -1$ at high redshifts, and $w$ increases
only slightly by low redshifts. These models depend on initial field 
values (in contrast with freezing  models of quintessence which depend on the shape of the potential).

In the present work we evaluate the cosmological consequences of the
evolving quintessence field in 
the context of thawing models by considering various observational datasets
and obtain plausible initial values of the scalar field $\phi_{i}$ . 
A lower bound on the the allowed values of $\phi_i$ 
has been previously obtained in Ref. \cite{Gupta:2011ip}.
Our current numerical analysis provides 
stronger constraints on $\phi_i$.
We further relate the initial value to
quantum fluctuations of $\phi$ during inflation and thereby to the duration of inflation.
The lower bound on $\phi_i$ then provides a lower bound on the number of e-foldings of inflation in our scenario where the initial
value of the quintessence field is determined by 
quantum fluctuations of the field
during inflation.
Our work in this article is organised as
follows.  In section \ref{thawing} we describe the thawing 
quintessence model in the standard minimal framework. In
section \ref{constraints} we provide a detailed description
of the datasets used to obtain the observational
constraints on the parameters of the model. In section \ref{analysis} we 
use the results obtained
in our investigation to obtain a lower bound on the initial 
value of $\phi$.  We then discuss the generation of the initial value by quantum
fluctuations during inflation and use the lower bound on $\phi_i$ to obtain a lower bound on the number
of e-foldings of inflation, $N$. We end with our conclusions in section
\ref{concl}.

\section{The Thawing Quintessence Scenario}\label{thawing}

We will assume that the dark energy is provided by a minimally coupled
scalar field $\phi$ with the equation of motion for the homogeneous 
component given by
\begin{equation}
\label{motionq}
\ddot{\phi}+ 3H\dot{\phi} + \frac{dV}{d\phi} =0 \,,
\end{equation}
where the Hubble parameter $H$ is given by
\begin{equation}
\label{H}
H = \left(\frac{\dot{a}}{a}\right) = \sqrt{\rho/3 M^{2}_{P}}\,.
\end{equation}
Here $a(t)$ is the scale factor, $\rho$ is the total density, and 
$M_{P}=(8 \pi G)^{-1/2}=2.4\times10^{18}\gev$ is the
reduced Planck mass.
Eq. (\ref{motionq}) indicates
that while the field rolls downhill in the potential $V(\phi)$,
its motion is damped by a term proportional to $H$.
The pressure and density of the
scalar field are given by
\begin{equation}
p = \frac{\dot \phi^2}{2} - V(\phi)
\end{equation}
and
\begin{equation}
\label{rhodense}
\rho = \frac{\dot \phi^2}{2} + V(\phi)
\end{equation}
respectively, and the equation of state parameter $w$
is given by Eq. (\ref{w}).
At late times, the Universe is dominated
by dark energy due to $\phi$,
and non-relativistic matter.
\footnote{In our numerical analysis we will go back in time till decoupling. At 
that epoch the energy density in radiation will be more than that in 
$\phi$.  The radiation component $\Omega_r$ is included in our numerical analysis.}
We assume a flat Universe
so that
$\Omega_\phi + \Omega_m + \Omega_r= 1$. Then Eqs. (\ref{motionq})
and (\ref{H}) can be rewritten in
terms of the variables $x$, $y$, and $\lambda$ defined by
\begin{eqnarray}
x & \equiv& \phi^\prime/(M_{P}\sqrt{6})\,, \\
y &\equiv& \sqrt{V(\phi)/(3H^2M^{2}_{P})}\,, \\
\lambda &\equiv& -\frac{M_{P}}{V}\frac{dV}{d\phi}\,,
\label{lambda}
\end{eqnarray}
where the prime denotes the derivative with respect to $\ln a$;
e.g., $\phi^\prime \equiv a(d\phi/da)$.
$x^2$ gives the contribution of the kinetic energy of the scalar
field to $\Omega_\phi$, and $y^2$ gives the contribution of the
potential energy, so that
\begin{equation}
\label{Om}
\Omega_\phi = x^2 + y^2\,,
\end{equation}
while the equation of state is rewritten as
\begin{equation}
\label{gamma}
\gamma \equiv 1+w = \frac{2x^2}{x^2 + y^2}\,.
\end{equation}
It is convenient to work in terms of $\gamma$, since we are
interested in models for which $w$ is near $-1$,
so $\gamma$ is near zero.
Eqs. (\ref{motionq}) and (\ref{H}),
in a Universe containing only matter and a scalar field,
become \cite{Scherrer:2007pu}
\begin{eqnarray}
x^\prime &=& -3x + \lambda\sqrt{\frac{3}{2}}y^2 + \frac{3}{2}x[1 + x^2-y^2]\,,
\label{xeqn}\\
y^\prime &=& -\lambda\sqrt{\frac{3}{2}}xy + \frac{3}{2}y[1+x^2-y^2]\,,
\label{yeqn}\\
\lambda^\prime &=& - \sqrt{6} \lambda^2(\Gamma - 1) x\,,
\label{lambdaeqn}
\end{eqnarray}
where
\begin{equation}
\label{Gamma}
\Gamma \equiv V \frac{d^2 V}{d\phi^2}/\left(\frac{dV}{d\phi} \right)^2\,.
\end{equation}

We numerically solve the system of Eqs. (\ref{xeqn}) - (\ref{lambdaeqn}),
initially for $V\sim\phi^{\pm2}$,
from an initial time $t_i$ at decoupling till $t_0$ today.
We choose $\gamma_i\sim0$,
and for different values of $\lambda_i$ and $\Omega_{m0}$
we obtain 
$x$, $y$
and $\lambda$ as functions of $a$.\footnote
{
Our code solves Eqs. 
(\ref{xeqn})-(\ref{lambdaeqn}) for initial values $x_i$, $y_i$ and $\lambda_i$,
where $x_i$ and $y_i$ are derived using Eqs. (\ref{Om}) and (\ref{gamma}) from initial values $\gamma_i$ and $\Omega_{\phi i}$.
We choose a small value of $\gamma_i\sim10^{-9}$ and do not vary it for our analysis.  
For a certain $\lambda_i$, we choose a value of $\Omega_{m0}$ and then vary $\Omega_{\phi i}$ till we
get a solution that satisfies $\Omega_{\phi0}=1-\Omega_{m0}-\Omega_{r0}$.
We identify this solution for $x$, $y$ and $\lambda$ as relevant for  
the chosen values of $\lambda_i$ and $\Omega_{m0}$
(having swapped the variable 
$\Omega_{\phi i}$ with $\Omega_{m0}$). 
This process is repeated for different values of $\lambda_i$ and
$\Omega_{m0}$. We also set $\Omega_{r0} = 9.22\times10^{-5}$ from 
Planck \cite{planck}.} 
The information about the
potential is encoded in the parameter $\Gamma$ given above.
We use our solutions for $x$ and $y$, 
and 
the relation $a^{-1}=1+z$ 
(with $a_0=1$), to obtain the normalized Hubble
parameter 
\begin{eqnarray}
&&\mathcal{H}^{2}(z,\lambda_{i},\Omega_{m0}) \equiv \frac{H^2(z,\lambda_{i},\Omega_{m0})}{H_{0}^2}\nonumber\\
&&= \frac{8\pi G}{3H_0^2}\left(\rho_{m0}(1+z)^3+\rho_{r0}(1+z)^4+\rho_\phi(z,\lambda_i,\Omega_{m0})\right)\nonumber\\
&&= \Omega_{m0}(1+z)^3+\Omega_{r0}(1+z)^4+\Omega_{\phi}(z,,\lambda_i,\Omega_{m0}) \mathcal{H}^2\nonumber\\
&&= \frac{\Omega_{m0}(1+z)^{3}+\Omega_{r0}(1+z)^{4}}{(1-\Omega_{\phi}(z,\lambda_{i},\Omega_{m0}))}\,,
\label{hubth}
\end{eqnarray}
where $\Omega_{\phi}$ is given by Eq. (\ref{Om}).

We then utilise this expression of the Hubble parameter to calculate 
the luminosity distance and angular diameter distance and relate them and $\mathcal{H}$ to the 
observations of type Ia supernovae (SN) data \cite{Suzuki:2011hu},
the baryon acoustic oscillations (BAO) data \cite{Giostri:2012ek}, the 
cosmic microwave background (CMB) shift parameter \cite{planck}
and 
the observational Hubble parameter (HUB) data \cite{Farooq:2013hq} and constrain the parameter space for 
each model defined by the values of $\lambda_{i}$ and $\Omega_{m0}$.
The allowed values of $\lambda_i$ from our numerical analysis give 
constraints on $\phi_i$, the value of $\phi$ at decoupling.
We then extend our analysis to study other power law potentials,  $V=A\phi^n$.
Thereafter, based on our arguments below, we set
$\phi_i\sim\phi_I$, where $\phi_I$ is
the value of $\phi$ at the end of inflation. We then study the 
conditions on inflation
to obtain $\phi_I$.

\section{Observational datasets}\label{constraints}

We use the $\chi^2$ analysis to constrain the parameters of our assumed parameterization. We will use the maximum likelihood method and obtain the total likelihood function 
for the parameters $\lambda_i$ and $\Omega_{m0}$ in a model
as the product of independent likelihood functions for each of the datasets being used.
The total likelihood function is defined as 
\begin{equation}
{\cal L}_{tot}(\lambda_i , \Omega_{m0}) \equiv e^{-\frac{\chi_{tot}^2(\lambda_i , \Omega_{m0})}{2}}\,,
\end{equation} 
 where
\begin{equation}
\chi_{tot}^2 = \chi_{SN}^2+ \chi_{BAO}^2+ \chi_{CMB}^2 +\chi_{HUB}^2
\label{chisqtot}
\end{equation}
is associated with the four datasets mentioned above.
The best fit value of parameters is obtained by minimising $\chi^2$  
with respect to 
$\lambda_i$ and $\Omega_{m0}$.
In a two dimensional parametric space, the likelihood contours in 1$\sigma$ and 2$\sigma$ confidence region
are given by $\Delta\chi^2 = \chi^2 - \chi_{min}^2 =$  2.3 and 6.17 respectively.

\subsection{Type Ia Supernovae}
Type Ia supernovae are very bright and can be observed at redshifts upto $z\sim 1.4$. They have nearly the same luminosity which is redshift independent and well calibrated by the light curves.
Hence they are very good standard candles.

The distance modulus of each supernova is defined as
\begin{equation}
\mu_{th}(z) = 5\log_{10} D^{th}_{L}(z)+\mu_{0}\,,
\end{equation}
where   the
theoretical
Hubble free luminosity distance $D_L^{th}$ in a flat universe for a model is given by

\begin{equation}
D_{L}^{th}(z,\lambda_i,\Omega_{m0}) = H_{0}d^{th}_{L} = (1+z)\int_{0}^{z}\dfrac{dz'}{\mathcal{H}
(z',\lambda_i,\Omega_{m0})}\,.
\label{DLth}
\end{equation}

Above, $d_L^{th}$ is the luminosity distance, $\mathcal{H}$

is
obtained from Eq. (\ref{hubth}), and
$\mu_0$ is the zero point offset. 

We  construct the $\chi^2$ for the supernovae analysis after marginalising
over the nuisance parameter $\mu_0$ as \cite{Nesseris:2005ur}
\begin{equation}
\chi^2_{SN}(\lambda_{i},\Omega_{m0}) = A(\lambda_{i},\Omega_{m0}) - \frac{B^2(\lambda_{i},\Omega_{m0})}{C}\,,
\label{chisqSN11}
\end{equation}
where 

\begin{eqnarray}
A(\lambda_{i},\Omega_{m0})&=& \sum_{i}\frac{\left(\mu_{obs}(z_i)-\mu_{th}(z_i,\mu_0=0,\lambda_{i},\Omega_{m0})\right)^2}{\sigma_{\mu_{obs}}^2(z_{i})}
\nonumber\\
B(\lambda_{i},\Omega_{m0})&=&\sum_{i}\frac{\left(\mu_{obs}(z_i)-\mu_{th}(z_i,\mu_0=0,\lambda_{i},\Omega_{m0}\right)}{\sigma_{\mu_{obs}}^2(z_{i})}\nonumber\\
C&=&\sum_{i}\frac{1}{\sigma_{\mu_{obs}}^2(z_{i})}\,.
\end{eqnarray}
($A$ should not be confused with the coefficient of the $V\sim\phi^{-2}$ potential.)
$\mu_{obs}(z_{i})$ is the observed distance modulus at a redshift $z_i$
and $\sigma_{\mu_{obs}}(z_{i})$ is the error in the measurement of $\mu_{obs}(z_{i})$. 
The latest Union2.1 compilation \cite{Suzuki:2011hu} of supernovae Type Ia data consists of the measurement of distance modulii $\mu_{obs}(z_{i})$ at 580 redshifts $z_{i}$ over the range $0.015\leq z \leq1.414$ with corresponding $\sigma_{\mu_{obs}}(z_{i}$). In our analysis
 we include the $\chi^2_{SN}$ given by Eq. (\ref{chisqSN11}) 
in Eq.  (\ref{chisqtot}).

\subsection{Baryon Acoustic Oscillations}

Baryon acoustic oscillations (BAO) refers to oscillations at sub-horizon length scales in the photon-baryon fluid
before decoupling due to collapsing baryonic matter and counteracting radiation pressure.
The acoustic oscillations  freeze at decoupling, and imprint their signatures on both the CMB (the acoustic peaks in the CMB 
angular power spectrum) and the matter distribution (the baryon acoustic oscillations in the galaxy power spectrum).

To obtain the constraints on our models from the BAO  
we use the comoving angular diameter distance
\begin{equation}
d_{A}^{th}(z_{*},\lambda_{i},\Omega_{m0}) = 
\frac{1}{H_0}
\int_{0}^{z_{*}} dz'/
\mathcal{H}(z',\lambda_{i},\Omega_{m0})\,.
\label{dAz}
\end{equation} 

The redshift at decoupling $z_{*}$ is obtained from the fitting formula in Ref. \cite{Hu:1995en}
as 1090.29. \footnote{In calculating
$z_{*}$ we have used the mean values of $\Omega_{b0}h^{2}$ and $\Omega_{c0}h^{2}$ from Planck+WP \cite{planck}.}
The dilation scale $D_V$ \cite{Eisenstein:2005su} is given by 
\begin{eqnarray}
D_{V}^{th} (z_{BAO},\lambda_i,\Omega_{m0})& = & H_0\left(\frac{z_{BAO}}{\mathcal{H}(z_{BAO},\lambda_{i},\Omega_{m0})}\right)^{1/3}\nonumber\\
& & \times \left(\int_0^{z_{BAO}}\frac{dz}{\mathcal{H}(z,\lambda_{i},\Omega_{m0})}\right)^{2/3}\nonumber
\end{eqnarray}
where $\mathcal{H}(z,\lambda_{i},\Omega_{m0})$ is
obtained from Eq. (\ref{hubth}). 
We construct the $\chi ^2$ for the BAO analysis as
\begin{equation}
\chi_{BAO}^2(\lambda_{i},\Omega_{m0}) = \sum_{ij}(x_{i}-d_{i})(C_{ij})^{-1}(x_{j}-d_{j})\,,
\end{equation}
where 
$x_{i} = \frac{d_{A}^{th}(z_{*},\lambda_{i},\Omega_{m0})}{D_{V}^{th}(z_{BAO},\lambda_{i},\Omega_{m0})}$ are 
the values predicted by a model, $d_{i}$ is the mean value of 
$\left(\dfrac{d_{A}^{obs}(z_{*})}{D_{V}^{obs}(z_{BAO})}\right)$ from observations given in 
Table 1 of Ref. \cite{Giostri:2012ek}, and $(C_{ij})^{-1}$
is the inverse covariance matrix in Eq. (3.24) of Ref. \cite{Giostri:2012ek}.
We include $\chi_{BAO}^2$ in Eq. (\ref{chisqtot}).

\subsection{Cosmic Microwave Background}
The locations of peaks and troughs of the acoustic oscillations in the CMB angular spectra are 
sensitive to the distance to the decoupling epoch. 
To obtain the constraints from CMB we utilise the 
\textit{shift parameter} $R$ from
Planck \cite{planck}. 
The CMB shift parameter $R$ is given by
\begin{equation}
R^{th}(z_{*},\lambda_{i},\Omega_{m0}) \equiv \sqrt{\Omega_{m0}H_{0}^{2}}\,d_{A}^{th}(z_{*},\lambda_{i},\Omega_{m0})
\end{equation}
where $d_{A}^{th}(z_{*},\lambda_{i},\Omega_{m0})$ is given by Eq. (\ref{dAz}).

The shift parameter $R(z_{*})$ from Planck observations is $1.7499\pm 0.0088$ \cite{planck}. 
Thus we obtain $\chi^2_{CMB}$ as
\begin{equation}
\chi_{CMB}^2(\lambda_{i},\Omega_{m0}) = \left(\frac{1.7499-R^{th}(z_{*},\lambda_{i},\Omega_{m0})}{0.0088}\right)^2\,.
\end{equation} 

\subsection{The observational Hubble parameter $H(z)$}

The parameter $H(z)$ describes the expansion history of the Universe and plays a central role in connecting dark energy theories and observations.
An independent approach, regarding the measurement of the expansion rate is provided by `cosmic clocks'. The best cosmic clocks are galaxies. 
The observational Hubble parameter data can be obtained based on 
differential ages of galaxies. 

Recently, Farooq et al. have compiled a set of 28 datapoints for $H(z)$ data, listed in Table 1 of Ref. \cite{Farooq:2013hq}. We use the measurements from Ref. \cite{Farooq:2013hq}
of the Hubble parameter $H_{obs}(z_{i})$ at redshifts 
$z_{i}$, with corresponding one standard deviation uncertainties $\sigma_{i}$, and
the current value of the Hubble parameter $H_0=67.3\pm1.2$ km $s^{-1} \rm{Mpc}^{-1}$ from 
Planck \cite{planck}.  Then for $\calH=H/H_0$
\begin{equation}
\chi^2_{HUB}
(\lambda_{i},\Omega_{m0}) = \sum_{i}
\left(\frac{\calH_{obs}(z_i)-\calH_{th}(z_i,\lambda_{i},\Omega_{m0})}{\sigma_{\calH_{obs}(z_i)}}\right)^2\,
\end{equation}
with $\calH_{th}(z,\lambda_{i},\Omega_{m0})$ obtained from  Eq. (\ref{hubth}), and 
$\sigma_{\calH_{obs}}$ obtained from $\sigma_{i}$ and the uncertainty in $H_0$.

\section{Results and Discussion}\label{analysis}

\begin{figure}[H]
\includegraphics[width=8cm]{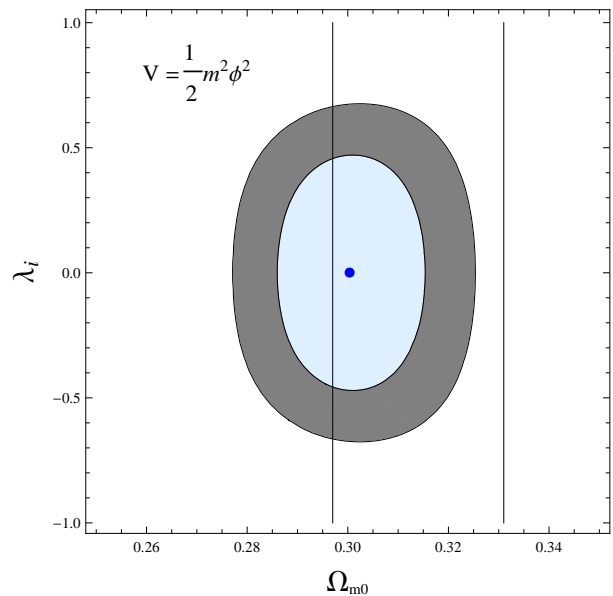}
\caption{The 1$\sigma$ and  2$\sigma$ confidence 
regions
in the 
$\lambda_{i}$ - $\Omega_{m0}$ plane for $V=m^2\phi^2/2$, or $ \Gamma = 0.5$, 
constrained by the SN+BAO+CMB+HUB data. The two vertical lines represent the Planck bounds on $\Omega_{m0}$. The thick dot represents the best fit values $\lambda_{i}=0.0008, \Omega_{m0}=0.3003$.}
\label{fig3}
\end{figure}

\begin{figure}[b]
\includegraphics[width=8cm]{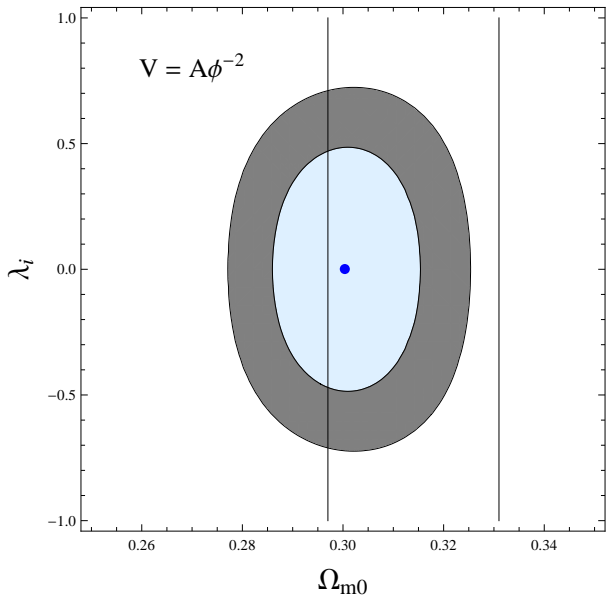}
\caption{The 1$\sigma$ and  2$\sigma$ confidence 
regions
in 
the $\lambda_{i}$ - $\Omega_{m0}$ plane for $V=A\phi^{-2}$, or $ \Gamma = 1.5 $, 
constrained by the SN+BAO+CMB+HUB data. The two vertical lines represent the Planck bounds on $\Omega_{m0}$. The thick dot represents the best fit values $\lambda_{i}=0.0009, \Omega_{m0}=0.3003$.}
\label{fig4}
\end{figure}

In Figs. \ref{fig3} and \ref{fig4} we present the likelihood contours  after combining all the datasets
for $V\sim \phi^{\pm2}$ .  
We find
that for $V\sim \phi^{2}$ the values of $|\lambda_i|$ greater than 0.67 are excluded
at 2 $\sigma$ confidence level whereas for $V\sim \phi^{-2}$ the values of $|\lambda_i|$ 
greater than 0.72 are excluded at 2 $\sigma$ confidence level. The best fit value for $\Omega_{m0}$ in both cases turns out to be 0.3003 which is within the bounds on $\Omega_{m0}$ by Planck \cite{planck}.  The best fit value for $\lambda_i$ is 0.0008 and 0.0009 respectively.

From Eq. (\ref{lambda}), 
\be
\lambda_{i}= \mp\frac{2 M_{P}}{\phi_i}
\ee

where $\mp$ refers to $V=\frac{1}{2} m^2 \phi^2,{A}{\phi^{-2}}$.
For $V=\frac{1}{2} m^2 \phi^2$, $|\lambda_{i}|< 0.67$
implies we need $|\phi_i| > 7.2\times10^{18}\gev$. 
For $V={A}{\phi^{-2}}$, $|\lambda_{i}| < 0.72$ implies we need $|\phi_i| > 6.7\times10^{18}\gev$. 
Hereafter we take 

\be \label{phiibound}|\phi_i|>7\times10^{18}\gev 
\ee 
for both potentials.
One may obtain a similar bound for the quadratic potential by equating
the energy density
in $\phi$ today, $\rho_{\phi0}=(1/2) m^2\phi_0^2\approx(1/2)\,m^2\phi_i^2$, with 
$\Omega_{\phi0} \, 3 H_0^2/(8\pi G)$, 
where one has assumed that the light quintessence field has not evolved much since decoupling.
Then for $\Omega_{\phi0}=0.7$ \cite{planck} and $m\lsim H_0$ one gets
\be
|\phi_i|\gsim5\times10^{18}\gev\,,
\label{estimatephii}
\ee
similar to the bound in Eq. (\ref{phiibound}).  
However, a priori one would not necessarily expect similar bounds. 
While using various observations to obtain the bound on $\phi_i$ in 
Eq. \eqref{phiibound}, we do not need to use that the mass of the quintessence 
field $m\leq H_{0}$.Given the current accuracy of observational data we obtain an upper bound on $\lambda_i$ of order 1, and hence a lower bound
for $\phi_i$ of order $M_{P}$, which agrees with the simple estimate above.  However if, for example,  more precise data
in the future implies a smaller
upper bound on $\lambda_i$ then the estimate of the lower bound for $\phi_i$ will rise and will not match with the bound in
Eq. \eqref{estimatephii}.  We also point out that the estimate above is only valid for a quadratic potential, while our approach
is more general.  Thus only an analysis as above taking into account various observations can give the relevant bounds 
on $\phi_i$.

\setlength{\textfloatsep}{5pt}

\setlength{\textfloatsep}{5pt}
We note that the likelihood contours for both the potentials look very similar. This indicates that the datasets that we compare
with are not able to distinguish between the different potentials under consideration. We have confirmed that the plots of $x, y, \lambda$ and $\Omega_\phi$ as functions
of $a$, for $\Omega_{m0}=0.3$ show variations between the potentials, but the key parameter that is 
relevant for comparison of the models with the different datasets is $\calH(z)$.

\begin{figure}[H]
\includegraphics[width=8.5cm]{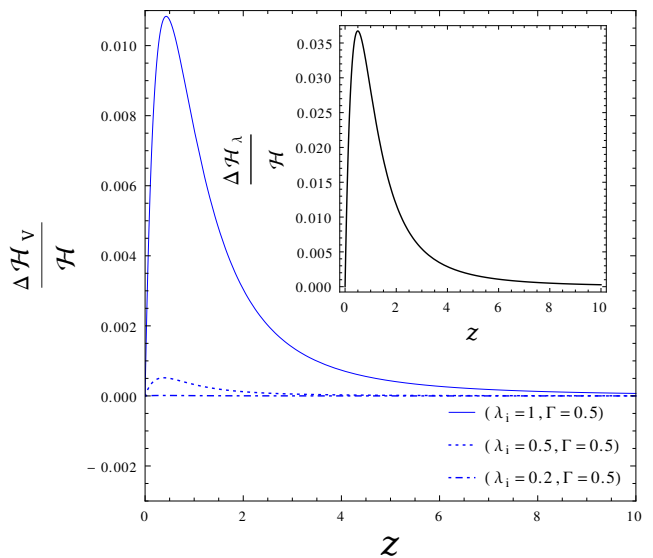}
\caption{The relative difference in the Hubble parameter,
$\frac{\Delta \mathcal{H}_V}{\mathcal{H}}$, as a function of redshift $z$
 for
  $V\sim\phi^{\pm2}$, with $\lambda_{i}$ = 0.2, 0.5 and 1 (from bottom to top) and $\Omega_{m0}=0.3$.
 The inset shows the relative difference in the Hubble parameter,
$\frac{\Delta \mathcal{H}_\lambda}{\mathcal{H}}$, as a function of redshift $z$
 for
   $\lambda_{i}$ = 1 and 0.2, for $V\sim\phi^{2}$ and $\Omega_{m0}=0.3$.
  }
\label{figH} 
\end{figure}

In Fig. \ref{figH} we plot $\frac{\Delta \calH_V}{\calH}$ against $z$,
where $\Delta \calH_V$ is the difference in $\calH(z)$ for the two potentials
$V\sim\phi^{\pm2}$ (the denominator corresponds to $V(\phi)\sim\phi^{2}$),  for different 
values of $\lambda_i$ with $\Omega_{m0} = 0.3$. The relative difference between the two 
potentials  for $\lambda_i=1$ reaches a maximum of about $1\%$ and is even smaller for
the other considered values of $\lambda_i$.  The small relative difference
is also seen in
Fig. 2 of Ref. \cite{SenSenSami2010}.  This is because in the
thawing scenario the field does not roll much in its potential and hence
is not sensitive to the form of the potential.

In the inset of Fig. \ref{figH} we plot $\frac{\Delta \calH_\lambda}{\calH}$ against $z$,
where 
$\Delta \calH_\lambda$ is the difference in $\calH(z)$ for $\lambda_i=1,0.2$
for $V\sim\phi^{2}$ (the denominator corresponds to $\lambda_i=0.2$),  
with $\Omega_{m0} = 0.3$.  Though the relative difference in $\calH(z)$ is small
for different values of $\lambda_i$,
the data does indicate a difference in the 
likelihoods for different points in the $\lambda_i-\Omega_{m0}$ plane.
This is because the SN and CMB datasets disfavor large values of $\lambda_i$. 
The SN data has a large
number of datapoints while the CMB shift parameter is very precisely measured and hence these datasets
are more sensitive to the variations in $\lambda_i$.

\begin{figure}[t]
\includegraphics[width=7cm]{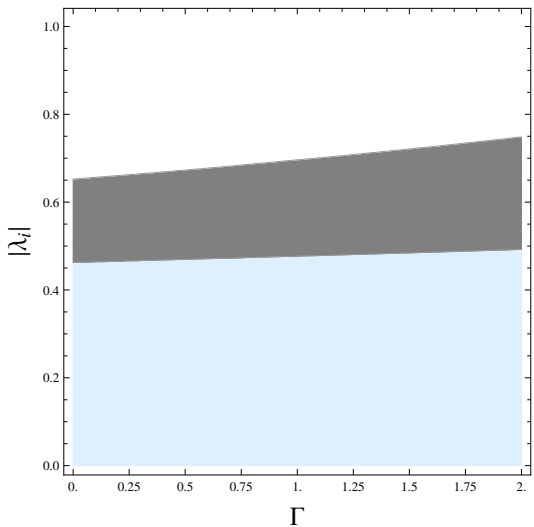}
\caption{The shaded region shows 1$\sigma$ (blue) and  2$\sigma$(grey) confidence regions in 
the $|\lambda_{i}|$ - $\Gamma$ plane 
constrained by the SN+BAO+CMB+HUB data. Discrete values for $\Gamma$ in steps of 0.25 are considered here. $\Gamma = 0, 0.5, 0.75, 1.25, 1.5, 2$ corresponds to $V\sim\phi,\phi^{2},\phi^4,\phi^{-4},\phi^{-2},\phi^{-1}$ respectively. 
We have taken $\Omega_{m0}=0.3$. In addition, we have also studied $\Gamma$ values of 2/3 
and 4/3, corresponding to 
$V\sim\phi^{\pm3}$.
}
\label{fig5}
\end{figure}

Thus far we have presented the results for $V(\phi)\sim\phi^{\pm2}$. In Fig. \ref{fig5} we present our results for bounds on $\lambda_i$ for different values of $\Gamma$ as
would be relevant for other potentials.  
Our analysis has been carried out
for discrete values of $\Gamma$ from 0 to 2, in intervals of 0.25.  
From Eq. (\ref{Gamma}), for potentials of the form
$V=A\phi^n$, where $A$ is taken as real and positive, 
$n=1/(1-\Gamma)$. Thus 
Fig. \ref{fig5} includes potentials with $n=\pm1,\pm2,\pm4$ for $\Gamma = 0, 2, 0.5, 1.5, 0.75, 1.25$ respectively.  In addition,
we have obtained the bounds for $n=\pm3$ corresponding to $\Gamma=2/3,4/3$ 
respectively.
For fractional $n$, the form of
the potential implies $\phi$ has
to be positive and so allowed values of $\lambda_i=-n M_P/\phi_i$ have the opposite sign of $n$. For integer values of $n$, if $n$ is
odd, positivity of the dark energy potential energy implies $\phi$ takes positive values.  Therefore we presume that $\phi$ takes positive
values and project the corresponding values of $\lambda_i$ in Fig. \ref{fig5}. To simplify the numerical analysis we have taken $\Omega_{m0}$ to be 0.3. The $1\sigma$ and $2\sigma$ bound on the parameter $\lambda_{i}$ is not very different from what we obtain in case of $V(\phi)\sim\phi^{\pm2}$.
Thus again $|\phi_i|>O(M_P)$.
 We mention that $\Gamma=1$ corresponds to the case of the exponential potential $V=B\exp({\pm b\,\phi/\phi_{i}})$ and the bound
 on $\lambda_i$ does not give a bound on $\phi_{i}$ due to the presence of the undetermined constant $b$. 

We now consider plausible values of $\phi_i$ that one may obtain in an inflationary
universe by considering a condensate of $\phi$ being generated due to quantum
fluctuations during inflation.
A light scalar field of mass $m$ ($m\ll H_I$, where $H_I$ is the 
Hubble parameter during inflation) will undergo 
quantum fluctuations during inflation.
The fluctuations are given by Eq. (7) of Ref. \cite{lindefluc}  
(we presume $H_I$ does not vary during inflation)

\be
\langle \delta\phi^2\rangle
=
\int_{a_i H_I}^{a H_I} \frac{d^3 k}{2\pi^3} |\delta \phi_k|^2
\\
=\frac{3 H_I^4}{8 \pi^2 m^2}
\left[1-\exp\left(-\frac{2m^2 N}{3 H_I^2}\right)\right]
\label{phisqtotal}
\ee

where $N$ is the number of e-foldings of inflation.
We consider the case where $m^2 N\ll H_I^2$. Then 
\be
\langle \delta\phi^2\rangle
=\left(\frac{H_I}{2\pi}\right)^2 N\,.
\label{phiisqN}
\ee 

Ignoring an initial value of $\phi$ at the beginning of inflation, and any potential driven classical evolution, the value
of $\phi$ at the end of inflation is 
\be\label{e-fold}
\phi_I =\pm\sqrt{\langle \delta\phi^2\rangle} = \pm\frac{H_I}{2\pi} \sqrt{N}
\ee

The above discussion is in the context of a quadratic potential, 
$V=\half m^2\phi^2$.  For
$V=A\phi^{-2}$, 
and $ A\phi^n$ considered in Fig. \ref{fig5},
we presume that the potential is
very flat during inflation, and so Eq.~(\ref{phiisqN}) again describes the 
evolution of $\phi$.

\begin{figure}[H]
\includegraphics[width=9cm]{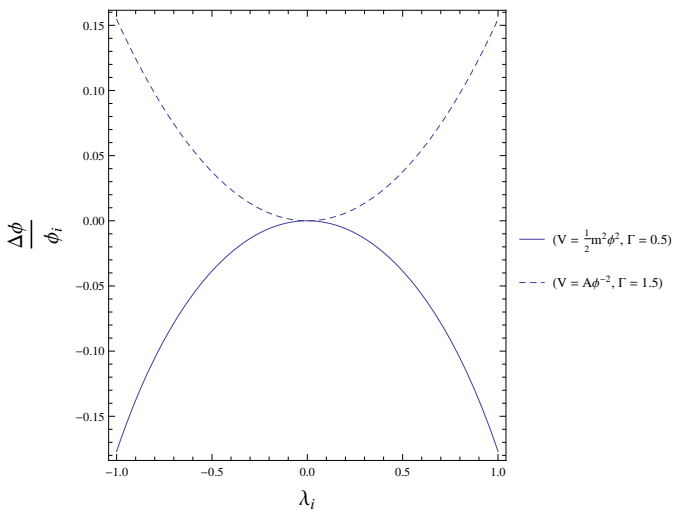}\caption{The 
fractional change in $\phi$ from decoupling to today as a function
of $\lambda_i$
 for $V=m^2\phi^2/2$ ($\Gamma = 0.5$, solid line) 
and  
$V=A\phi^{-2}$ ($\Gamma = 1.5$, dashed line). 
We have taken $\Omega_{m0}=0.3$ here.
For $|\lambda_i|<0.7,$ the fractional change is less than $0.1$, or $10\%$
\label{fig2}}
\end{figure}

After inflation $\phi$ will evolve in two ways.
\begin{itemize}
\item $\phi$ will evolve classically in its potential $V(\phi)$.  Our numerical analysis
in Fig. \ref{fig2} shows that the field does not evolve much between decoupling
and the present epoch (less than 10 \% for $|\lambda_i|<0.7$). We expect its evolution is slower at earlier times. Therefore we may ignore the 
evolution of $\phi$ from the end of inflation till decoupling. (For
a quadratic potential, $m\ll H_0\ll H_I$ and therefore the assumption $\phi_i \sim \phi_I$ is 
obviously justified.)
\item
After inflation, modes of the $\phi$ field re-enter the horizon 
(during the radiation and matter dominated eras).
These should no longer be considered part of the homogeneous $\phi$ 
condensate at late time. This can affect $\phi$ by removing $\phi$ fluctuations
generated over the last 30-60 e-foldings of (electroweak to GUT scale) inflation. But our constraint on $N$ below will be many orders of 
magnitude larger so we can ignore this too in our use of Eq. (\ref{e-fold}).

\end{itemize}
The above discussion implies that we may take $\phi_i\approx\phi_I$, i.e., we may assume that the field $\phi$ has not evolved
much from inflation till decoupling.
From our analysis above we have $|\phi_i|>7\times10^{18}\gev$ for $V\sim\phi^{\pm2}$.  Then, from Eq. (\ref{e-fold}),
we get a lower bound on $N$ as
\begin{equation}
N>\frac{2\times 10^{39}\gev^2}{H_I^2}
\label{Nbound1}
\end{equation}

From the Planck bound on the tensor-to-scalar ratio, $H_I<9\times10^{13}\gev$ \cite{planck-inflation}.
Therefore we finally obtain
\begin{equation}
N > 2\times 10^{11}\,.
\label{Nbound2}
\end{equation}
A similar bound will apply to the other power law potentials $V\sim\phi^n$.

Combining Eq. (\ref{Nbound1}) and the Planck bound on $H_I$ we get $H_I^2/N<3\times10^{16}\gev$.
For a quadratic quintessence potential with $m\lsim H_0
\approx 10^{-42}\gev$ one sees that our assumption that $m^2 \ll H_I^2/N$ is highly feasible.

Ref. \cite{ringevaletal} also considers a dark energy condensate from quantum fluctuations of the quintessence
field during inflation.  Their analysis is primarily in the asymptotic limit  $m^2 N\gg H_I^2$ of Eq. (\ref{phisqtotal}).
They also consider $m^2 N\ll H_I^2$ with an argument similar to that in the paragraph following Eq. (\ref{phiibound}).

\section{Conclusions}\label{concl}

In this article we have first numerically solved for the evolution of the quintessence field $\phi$ in thawing models of dark energy
for potentials of the form
$V\sim\phi^{\pm2}$
and different values of $\Omega_{m0}$ and 
$\lambda_i=-(M_P/V)dV/d\phi=\mp M_{P}/\phi_i$, where
$\phi_i$ is the field value at decoupling.  From this we obtain the Hubble parameter $H(z)$, and the luminosity distance and angular diameter distance, as given in Eqs. (\ref{hubth}, \ref{DLth}) and (\ref{dAz}) respectively.  We then relate them to the 
observations of type Ia supernovae data,
the baryon acoustic oscillations data, the 
cosmic microwave background shift parameter and 
the observational Hubble parameter data  to constrain the values of $\lambda_{i}$ and $\Omega_{m0}$.
The likelihood contours for the two potentials looks very similar (due to similar $H(z)$ behaviour)
and in both the cases we obtain $|\phi_i| > 7 \times 10^{18}\gev$. We extend the analysis to other potentials of the form
$V=A\phi^n$ and obtain the bound on 
the initial value of the quintessence field, 
assuming $\Omega_{m0} = 0.3$. 
Once again, we find that 
$|\phi_{i}| > O(M_{P})$. We have then argued that $\phi$ does not
evolve much between the end of inflation and decoupling and further
considered a scenario where
the field value at the end of inflation $\phi_I$ is due to quantum fluctuations of $\phi$ during inflation.  This allows
us to use the lower bound on $\phi_i\approx\phi_I$ to constrain the duration of inflation -- the number of e-foldings $N$
is constrained to be greater than
$2\times10^{11}$, for $H_I<9\times10^{13}\gev$.

The inflationary paradigm does not stipulate an upper bound on $N$ and so large values of $N$ such as that required above
are plausible.  
Large values of $N$ are more likely in large field inflation models as
discussed in Ref. \cite{remmen_carroll}. 
However very large values of $N$ can imply a large variation $(\gg M_P)$
in the inflaton field
 during inflation which can be problematic if the inflationary scenario
constitutes a low energy effective field theory derivable from some
 Planck scale theory \cite{baumannmcallister}.  
On the other hand, one can get a large value of $N$
in eternal inflation scenarios without a large net variation in the inflaton field \cite{Guth:2007ng}.

\begin{acknowledgements}

RR would like to acknowledge Gaurav Goswami for useful discussions. We
thank Anjishnu Sarkar for help  with the plots. We thank an anonymous referee for valuable suggestions on improving this work.

\end{acknowledgements}


\begin{thebibliography}{99}

\bibitem{Riess:1998cb} 
  A.~G.~Riess {\it et al.}  [Supernova Search Team Collaboration],
  Astron.\ J.\  {\bf 116}, 1009 (1998)
  [astro-ph/9805201].
\bibitem{Perlmutter:1998np} 
  S.~Perlmutter {\it et al.}  [Supernova Cosmology Project Collaboration],
  Astrophys.\ J.\  {\bf 517}, 565 (1999)
  [astro-ph/9812133].
  
\bibitem{Tonry:2003zg} 
  J.~L.~Tonry {\it et al.}  [Supernova Search Team Collaboration],
  Astrophys.\ J.\  {\bf 594}, 1 (2003)
  [astro-ph/0305008].
  
\bibitem{Zlatev:1998tr} 
  I.~Zlatev, L.~M.~Wang and P.~J.~Steinhardt,
  Phys.\ Rev.\ Lett.\  {\bf 82}, 896 (1999)
  [astro-ph/9807002].

\bibitem{Copeland:2006wr} 
  E.~J.~Copeland, M.~Sami and S.~Tsujikawa,
  Int.\ J.\ Mod.\ Phys.\ D {\bf 15}, 1753 (2006)
  [hep-th/0603057].

\bibitem{Li:2011sd} 
  M.~Li, X.~D.~Li, S.~Wang and Y.~Wang,
  Commun.\ Theor.\ Phys.\  {\bf 56}, 525 (2011)
  [arXiv:1103.5870].
  
\bibitem{Caldwell:2005tm} 
  R.~R.~Caldwell and E.~V.~Linder,
  Phys.\ Rev.\ Lett.\  {\bf 95}, 141301 (2005)
  [astro-ph/0505494].

\bibitem{Gupta:2011ip} 
  G.~Gupta, S.~Majumdar and A.~A.~Sen,
  Mon.\ Not.\ Roy.\ Astron.\ Soc.\  {\bf 420}, 1309 (2012)
  [arXiv:1109.4112] 

\bibitem{Scherrer:2007pu} 
  R.~J.~Scherrer and A.~A.~Sen,
  Phys.\ Rev.\ D {\bf 77}, 083515 (2008)
  [arXiv:0712.3450].

\bibitem{Suzuki:2011hu} 
  N.~Suzuki, D.~Rubin, C.~Lidman, G.~Aldering, R.~Amanullah, K.~Barbary, L.~F.~Barrientos and J.~Botyanszki {\it et al.},
  Astrophys.\ J.\  {\bf 746}, 85 (2012)
  [arXiv:1105.3470].
  
\bibitem{Giostri:2012ek} 
  R.~Giostri, M.~V.~d.~Santos, I.~Waga, R.~R.~R.~Reis, M.~O.~Calvao and B.~L.~Lago,
  JCAP {\bf 1203}, 027 (2012)
  [arXiv:1203.3213].   
  
\bibitem{planck}
  P.~A.~R.~Ade {et al.}  [Planck Collaboration],
  Astron.\ Astrophys.\  (2014)
  [arXiv:1303.5076].

\bibitem{Farooq:2013hq} 
  O.~Farooq and B.~Ratra,
  Astrophys.\ J.\  {\bf 766}, L7 (2013)
  [arXiv:1301.5243].
  
  
\bibitem{Nesseris:2005ur} 
  S.~Nesseris and L.~Perivolaropoulos,
  Phys.\ Rev.\ D {\bf 72}, 123519 (2005)
  [astro-ph/0511040].
  
  
\bibitem{Hu:1995en} 
  W.~Hu and N.~Sugiyama,
  Astrophys.\ J.\  {\bf 471}, 542 (1996)
  [astro-ph/9510117].

\bibitem{Eisenstein:2005su} 
  D.~J.~Eisenstein {\it et al.}  [SDSS Collaboration],
  Astrophys.\ J.\  {\bf 633}, 560 (2005)
  [astro-ph/0501171].
  
\bibitem{SenSenSami2010}  
S. Sen, A. A. Sen and M. Sami, 
Phys.\ Lett.\ B {\bf 686}, 1 (2010). 

\bibitem{lindefluc}
A.~D.~Linde, 
Phys.\ Lett.\ B {\bf 116}, 335 (1982). 


\bibitem{planck-inflation} 
P.~A.~R.~Ade {\it et al.}  [Planck Collaboration],
  Astron.\ Astrophys.\  {\bf 571}, A22 (2014)
  [arXiv:1303.5082 [astro-ph.CO]].
 

\bibitem{ringevaletal}
  C.~Ringeval, T.~Suyama, T.~Takahashi, M.~Yamaguchi and S.~Yokoyama,
  Phys.\ Rev.\ Lett.\  {\bf 105}, 121301 (2010)
  [arXiv:1006.0368].
  

\bibitem{remmen_carroll} 
  G.~N.~Remmen and S.~M.~Carroll,
  Phys.\ Rev.\ D {\bf 90}, 063517 (2014)
  [arXiv:1405.5538].  
  
\bibitem{baumannmcallister}
  D.~Baumann and L.~McAllister,
  arXiv:1404.2601 [hep-th], see Sec. 4.3.
  

\bibitem{Guth:2007ng} 
  A.~H.~Guth,
  J.\ Phys.\ A {\bf 40}, 6811 (2007)
  [hep-th/0702178].  

\end{thebibliography}
\end{document}